# Effects on Amorphous Silicon Photovoltaic Performance from High-temperature Annealing Pulses in Photovoltaic Thermal Hybrid Devices

M.J.M. Pathak[a], J.M. Pearce[a,b*] and, S.J. Harrison[a]

[a] Department of Mechanical and Materials Engineering, Queen's University, Kingston, ON, Canada

[b] Department of Materials Science & Engineering and Department of Electrical & Computer Engineering, Michigan Technological University, Houghton, MI, USA

* Contact author: 601 M&M Building, 1400 Townsend Drive, Houghton, MI 49931-1295
ph: 906-487-1466 fax: 906-487-2630 email: pearce@mtu.edu

**Abstract**

There is a renewed interest in photovoltaic solar thermal (PVT) hybrid systems, which harvest solar energy for heat and electricity. Typically, a main focus of a PVT system is to cool the photovoltaic (PV) cells to improve the electrical performance, however, this causes the thermal component to under-perform compared to a solar thermal collector. The low temperature coefficients of amorphous silicon (a-Si:H) allow for the PV cells to be operated at higher temperatures and are a potential candidate for a more symbiotic PVT system. The fundamental challenge of a-Si:H PV is light-induced degradation known as the Staebler-Wronski effect (SWE). Fortunately, SWE is reversible and the a-Si:H PV efficiency can be returned to its initial state if the cell is annealed. Thus an opportunity exists to deposit a-Si:H directly on the solar thermal absorber plate where the cells could reach the high temperatures required for annealing.

In this study, this opportunity is explored experimentally. First a-Si:H PV cells were annealed for 1 hour at 100˚C on a 12 hour cycle and for the remaining time the cells were degraded at 50˚C in order to simulate stagnation of a PVT system for 1 hour once a day. It was found that, when comparing the cells after stabilization at normal 50˚C degradation, this annealing sequence resulted in a 10.6% energy gain when compared to a cell that was only degraded at 50˚C.

## 1. Introduction

Photovoltaic solar thermal (PVT) hybrid systems have been shown to be more efficient at solar energy collection on the basis of exergy, energy and cost [1-7]. Most current PVT systems are based on crystalline silicon (c-Si) photovoltaic (PV) materials, whose performance declines with increasing temperature by 0.04%/˚C [8]. Thus, these PVT system designs are primarily focused on cooling the c-Si PV cells to maximize the electrical gain, while the extracted thermal energy is viewed as a secondary benefit [7]. The result of this design focus is that the thermal component of the PVT system significantly under-performs when compared to standard solar thermal collectors [9-12]. Focusing the design on only maximizing solar electrical output thus prevents the PVT system from being optimized.

A potential solution to this challenge is to use a PV material that can operate at higher temperatures without substantial performance losses. One such promising material is hydrogenated amorphous silicon (a-Si:H), which has a temperature coefficient of only -0.01%/ºC [8], a quarter that of c-Si. A fundamental challenge recognized shortly after the discovery of the a-Si:H PV cell, was light-induced degradation of the a-Si:H PV performance known as the Staebler-Wronski effect (SWE) [13-17]. SWE causes defect states to form in the a-Si:H material when exposed to sunlight, which lower the efficiency of the PV cell. For commercial high-quality a-Si:H materials, these defects saturate after a long (~100 hours of continuous 1 sun) illumination and the efficiency of the cell is considered stabilized at this point [18-20]. This is stabilized state is known as the degraded steady-state (DSS). Fortunately, SWE is reversible and the a-Si:H PV cell efficiency can be returned to its initial state if the cell is heated to 150ºC for four hours as the defect states are annealed [13, 21-23] . Various schemes have been devised to take advantage of thermal annealing in a-Si:H, such as removing the entire PV array and annealing them in an air oven at lower temperatures (e.g. at 80ºC) over extended times [24]. In addition, it has been found that a-Si:H PV performs better at higher temperatures since the optoelectronic properties of a-Si:H materials [25-27] stabilize at a higher efficiency [18,28]. In a solar thermal flat plate collector the temperature can easily reach over 100ºC and even climb higher than 200ºC if the system is stagnated [29]. Thus an opportunity exists to deposit a-Si:H directly on the solar thermal absorber plate [30,31] where the cells could reach the high temperatures required for annealing. This suggests that a-Si:H would be an excellent choice of PV material for a PVT system and that careful control of the temperature of the thermal side of the PVT could be used to introduce thermal annealing pulse cycles to raise the overall PV electrical conversion performance.

This paper explores this potential and reports on a series of experiments in which a-Si:H based PV cells were exposed to high temperature pulse annealing cycles for which the temperature and duration were determined by the potential stagnation conditions in a PVT system. The results are discussed to determine the value of utilizing a scheme of thermal cycling in PVT systems to maximize the overall solar energy harvested.

**2. Materials and Methods**

The a-Si:H PV cells used in this study were fabricated by an AKT plasma enhanced chemical vapour deposition (PECVD) chamber. The i-layer thicknesses of 210, 420, 630 and 840 nm were deposited by ThinSilicon in the following bottom up structure as can be seen in Figure 1: AGC float glass (3mm)/SnO$_2$:F (700 nm) - textured fluorinated tin oxide/ Ag (200nm)/ AZO (100nm)/ n-a-Si:H (25nm)/ i-a:Si:H (210nm to 840 nm)/ p-a-Si:H (15nm)/ ITO (70nm).

*{Insert Figure 1 Here}*

The following sections describe the various methodologies required to complete and acquire the data for this study.

## 2.1 Light-Induced Degradation

A Newport class AAA solar simulator was used to apply AM1.5 1 sun illumination to the PV cells until a DSS was reached. Degradation temperatures were maintained with a Chemat Technology Inc. hot plate attached to a Cole Parmer Digi-Sense k-Type temperature controller. A k-type thermocouple was placed beside the cell to monitor and maintain surface temperature at the desired temperatures of 25°C (standard testing conditions for PV), 50°C (representative operating temperature for PV) or 90°C (representative of operating temperatures for solar thermal systems) during the degradation. PV Measurements I-V Curve software K2400 I-V was used with a Keithley 2000 multimeter and a Keithley 2400 source meter to measure the current-voltage output of the cells. An AutoIt macro was implemented to run the program to take the measurements at desired intervals.

## 2.2 Spike Annealing at the DSS

After reaching the DSS, the cells were exposed to a thermal annealing cycle (spike annealing) with a set point of 100°C for one hour. This test was to simulate stagnation of the thermal component of the PVT system. The 100 °C was determined as feasible by experimentally simulating stagnation using a test rig containing the PV cell array used in this study. It should be noted that this value is conservative due to the nature of the areas around the test cells, which were highly reflective. The 1 hour annealing time was considered a short enough period of time to not greatly affect the thermal system performance, but long enough to ensure a substantial annealing of the SWE defect states in the PV cell. The PV cell was then allowed to cool to its degraded temperatures of 90, 50 and 25°C, respectively and remained at these temperatures for at least 10 minutes to obtain an accurate temperature correlation.

## 2.3 Spike Annealing Cycle

Following the methods outlined in Section 2.1 and 2.2, a spike annealing cycle test was completed on a 12 hour cycle. The cell was degraded at 50°C for 10 hours and 45 minutes, at which time the surface temperature of the cell was raised to 100°C over approximately 15 minutes and was maintained at that temperature for 1 hour and then allowed to cool naturally to 50°C. It should be noted that the typical operating temperature for a PV module is 50°C [32].

## 3. Results and Discussion

It is known that a-Si:H PV when degraded at higher temperatures will stabilize at higher efficiencies, as can be seen in Figure 2 showing the normalized maximum power ($P_{max}$) with a 630 nm i-layer thick a-Si:H cell degraded at 25, 50 and 90°C [19].

*{Insert Figure 2 Here}*

From Figure 2, it can be easily determined that running an a-Si:H cell at higher temperatures will increase the energy output of the cell due to the earlier occurrence of the DSS. This supports the concept of using a-Si:H in PVT systems as a-Si:H cells stabilize at higher efficiencies at higher operating temperatures. However, it would be more beneficial to the overall energy conversion if the cell could retain its initial efficiency rather than operate at the DSS lower efficiency. In order to test if this was possible, high-temperature annealing pulses (or spike annealing) were investigated. The spike annealing process was applied to the stabilized cells of the temperature series found in Figure 2 and the results are shown for degradation temperatures of 25, 50 and 90°C, respectively, in Figures 3-5. In each graph, the fill factor (FF), maximum power and annealing temperature profile are plotted as a function of time. The FF indicates the quality of the cell performance as the ratio of actual power output to theoretical power output .The annealing temperature profile displays the change in temperature over time applied to the cell. It should be noted that the end points of each line segment are known, but changes between them are not displayed for clarity.

*{Insert Figure 3 Here}*

*{Insert Figure 4 Here}*

*{Insert Figure 5 Here}*

As can be seen in Figures 3-5, there are some interesting patterns that arise at all degradation temperatures investigated. During the ramp up in temperature, the power drops initially, but then slowly increases thereafter. This may be because the cell initially suffers from the rapid increase in the temperature during the ramp until it is closer to achieving surface cell temperatures of 100 °C required for the annealing process to take a significant effect [24]. Although a-Si:H PV do perform better at higher temperatures to a point, cells are also very sensitive to fluctuations in temperature. This can clearly be seen in Figures 4-5 at the 100 °C plateau, where the temperature fluctuated by a couple degrees resulting in a wavelike scattered pattern during this period. In all three annealing tests, the FF spiked at around 80 °C whereas the power reaches its maximum at temperatures lower than 50 °C. These findings can be explained by understanding the temperature relationships between the short circuit current ($I_{sc}$) and open circuit voltage ($V_{oc}$). $I_{sc}$ increases with temperature slightly, whereas the $V_{oc}$ decreases significantly with temperature, typically 3-4 times as fast. Therefore, at 80 °C the conditions are set such that the $I_{sc}$ has only dropped a little while the $V_{oc}$ has increased significantly with the result being a greatly increased FF. At 50 °C, the $V_{oc}$ has increased dramatically allowing for a larger maximum voltage and therefore greatly increasing the $P_{max}$ output.

Reconsidering Figure 2, it is clear that the starting points for cycles shown in Figures 3-5 are dependent on what temperature the cells are degraded. As mentioned before, the higher the temperature, the faster the DSS and higher the corresponding $P_{max}$. At the lower degraded temperatures in Figures 3-4, the annealing has a larger effect on the power increase compared to the higher temperature in Figure 5. This is expected given that the DSS $P_{max}$ is highest at 90 °C, such that the power gain is smaller when being annealed at increase in temperature of only 10 degrees. To compare the relative regeneration of the cells, the final $P_{max}$ in Figures 3-5 were divided by the initial $P_{max}$ in Figure 2, which resulted in percentage regenerations of 84.1%, 86.2% and 96.8% for the 25 °C, 50 °C and 90 °C degradations respectively. Thus, at the highest degradation temperature, the cell is almost fully regenerated in the spike annealing, even though the gain is the smallest. This can be understood again by noting that the cell stabilizes at a higher $P_{max}$ at higher temperature.

The degradation at 50 °C and relevant spike annealing cycles were completed for different i-layer thicknesses (i.e., 210, 420, 630 and 840 nm) to determine if there was a potential to use spike annealing to enable thicker cells (with higher efficiencies) than those currently used in industry. It was found that the patterns of response were consistent for all PV cells with different i-layer thicknesses as seen in Figure 6. Figure 6 shows the maximum power of 4 cell thickness PV devices as a function of time. The different shades of the symbols indicate the different temperature stages of the pulse annealing test: i) ramp up, ii) 100 °C plateau, iii) cooling down, iv) 50 °C plateau and v) cooling down again. The darker regions indicate the temperature plateaus of the spike annealing.

*{Insert Figure 6 Here}*

For the thinner the cell, the effect of a one hour spike anneal on the power of the cell is greatest. This is by virtue of the thicker cells having more material, which means they have more defect states that require a greater annealing time [33, 34]. Therefore, after the 1 hour of annealing, the thicker cells will have more defects remaining, which negatively impact their relative performance. It is clear that all a-Si:H PV devices, regardless of thickness between 200nm and 800nm i-layers benefited from the spike annealing cycle.

A realistic application of the spike annealing concept would be to apply it once a day in a PVT system using stagnation for a short time period (1 hour). This is necessary to minimize the detrimental impact on the thermal component of the system (e.g. any thermal energy used to anneal the PV is not collected as useful thermal energy). This daily thermal annealing pulse sequence was simulated and the results are shown in Figure 7 for a 630 nm i-layer thick PV cell. Figure 7 compares the same cell degraded at 50 °C with and without spike annealed at 100 °C in a 12 hour cycle period for a duration of just over 1 week described in Section 2.3.

*{Insert Figure 7 Here}*

From Figure 7, the comparison between a normal 50 °C degradation to a 12 hour degradation and spike annealing cycle reveals that, over the course of just over 1 week, the spike anneal cycle test produces 8.5% more energy if both were run for 192 hours. When comparing the two tests from the point where the normal 50 °C degradation test stabilized (140 hours) to the end of the test (192 hours), it was found that the spike annealing cycle test produced 10.6% more energy. Since there is very little change in the spike annealing cycle test from 140 hours onwards, this would imply that over the course of the lifetime of the PVT system, applying the spike annealing once a day would produce 10.6% more energy than if the system was not annealed at all. If the a-Si:H cell was 10% efficient this would mean the cell's overall effective efficiency would increase to just over 11%. These results are extremely promising as the improved performance of a-S:H PV in PVT systems using a spike annealing sequence would substantially improve the levelized cost of electricity from the devices [35].

## 4. Future work

In this study, relatively low-efficiency single junction a-Si:H devices were investigated, however, the use of a micromorph or multi-junction a-Si:H-based cells may increase the impact of thermal annealing cycles even further because of their overall improved performance [36]. In addition, it is suggested that an i-layer thickness series study be completed utilizing spike annealing cycles at higher temperatures for a shorter period of time might also produce the same effects, but have less of a negative effect on the thermal component. Future work is needed to assess the feasibility of stagnating the PVT system to determine if the gain in exergy from the solar PV cells exceeds the losses in thermal energy output due to the stagnation period. It would also be interesting to determine the best time during a day to implement the stagnation, which would produce the required annealing temperature while minimizing the stagnation thermal losses. For example, in the middle of the day where the irradiance is highest, a high stagnation temperature can be achieved but the thermal losses would be greater than if this was completed at the end of the day.

Finally, in recent years the use of Feed-In-Tariffs (FIT) programs has been introduced to various countries to promote green technologies. In Ontario, Canada, the FIT program offers a substantial financial benefit for each kWh of solar electricity provided to the grid [37, 38]. Future work is needed to investigate both the optimal dispatch strategy of thermal annealing in PVT devices to maximize income under different FIT structures, but also can be used to encourage energy policy formation that maximizes exergy for a given technology. For example, it may therefore be more beneficial from an economic perspective to spike anneal even in the middle of the day since the electricity gained greatly exceeds the cost of purchasing it even though from an exergy or greenhouse gas mitigation standpoint later thermal cycling may be optimal. Another concept that can be explored is utilizing dispatch strategies that only allow spike annealing at times when the thermal demand is zero or when the occupants will not be using the hot water for a while, such as when the occupants of a home are on vacation. This would then have no negative impact on the performance of the thermal component while improving the electrical efficiency.

## 5. Conclusions

This paper explored the potential of thermal annealing cycles (spike annealing) of a-Si:H based PV cells by simulating high temperature sequences that would occur during imposed stagnation conditions in hybrid PVT solar collectors. Through simulating the stagnation of an a-Si:H PVT system for 1 hour where the cell was maintained at 100 ˚C, it was found that there was 10.6% gain in solar electricity production compared to running a cell only at 50 ˚C once both had reached the normal 50˚C stabilized state. Assuming the rates of degradation were maintained throughout their lifetime, the relative gain would be maintained for the cells with spike annealing. With the prospects of producing substantially more energy due to spike annealing, PVT systems have the technical potential to become a much larger fraction of the burgeoning solar market.

## Acknowledgements

This work was supported by the Natural Sciences and Engineering Research Council of Canada, Canada Foundation for Innovation, Ministry of Research and Innovation, the Canadian

Solar Building Network, and PV Measurements Inc. The authors would like to acknowledge K. Girotra for providing samples.

<>Published as: M.J.M. Pathak, J.M. Pearce, and S.J. Harrison, "Effects on Amorphous Silicon Photovoltaic Performance from High-temperature Annealing Pulses in Photovoltaic Thermal Hybrid Devices" *Solar Energy Materials and Solar Cells*, 100, 199-203 (2012).
http://dx.doi.org/10.1016/j.solmat.2012.01.015

**Figures**

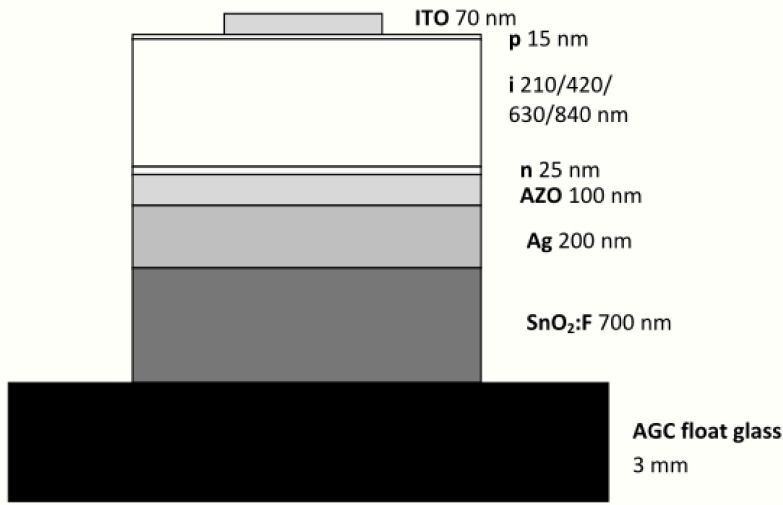

**Figure 1:** The composition and structure of the a-Si:H solar photovoltaic cells used in this study.

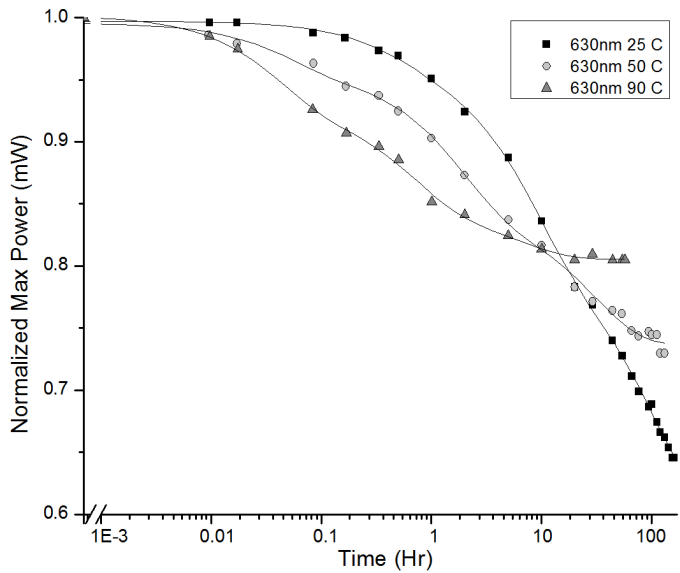

**Figure 2:** A normalized temperature series of 25, 50 and 90 °C degradation under 1 sun for a PV cell with an i-layer thickness of 630 nm.

<>Published as: M.J.M. Pathak, J.M. Pearce and, S.J. Harrison, "Effects on Amorphous Silicon Photovoltaic Performance from High-temperature Annealing Pulses in Photovoltaic Thermal Hybrid Devices" *Solar Energy Materials and Solar Cells*, 100, 199-203 (2012).
http://dx.doi.org/10.1016/j.solmat.2012.01.015

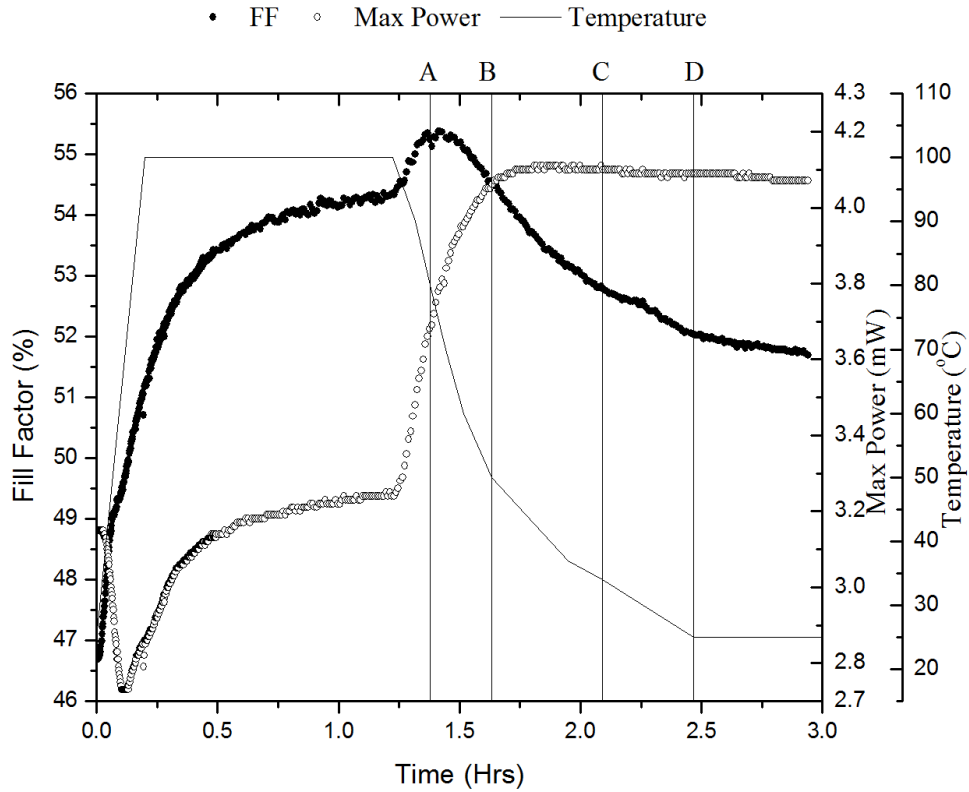

**Figure 3:** High-temperature annealing pulses (spike annealing) results for 25 °C degradation showing the fill factor, maximum power and temperature (A: 80 °C, B: 50 °C, C: fan turned on, D: 25 °C) as a function of time.

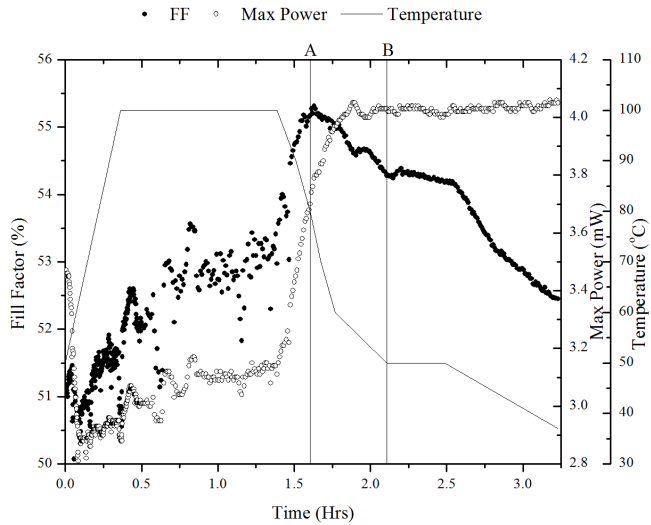

**Figure 4**: High-temperature annealing pulses (spike annealing) results for 50 °C degradation showing the fill factor, maximum power and temperature (A: 80 °C, B: 50 °C) as a function of time.

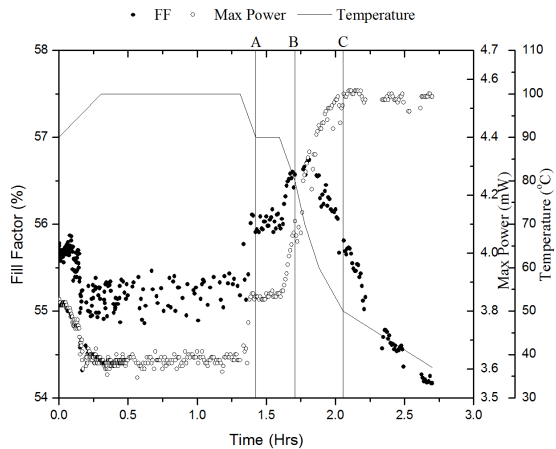

**Figure 5:** High-temperature annealing pulses (spike annealing) results for 90 °C degradation showing the fill factor, maximum power and temperature (A: 90 °C, B: 80 °C, C: 50 °C) as a function of time.

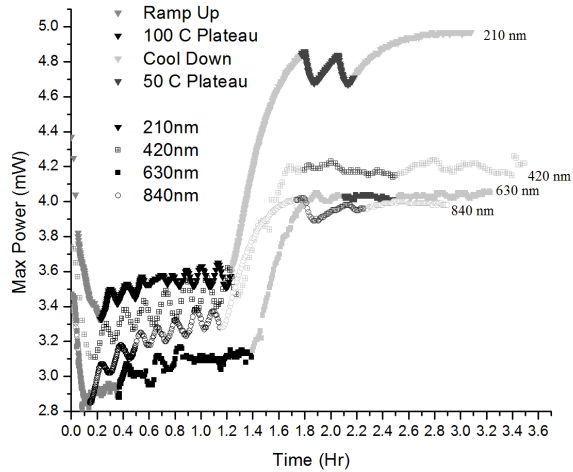

**Figure 6:** High-temperature annealing pulses (spike annealing) thickness series showing the maximum power as a function of time using a-Si:H cells with i-layer thicknesses of 210, 420, 630 and 840nm degraded at 50 °C under 1 sun. The color of the symbols indicates the stage of the temperature cycle.

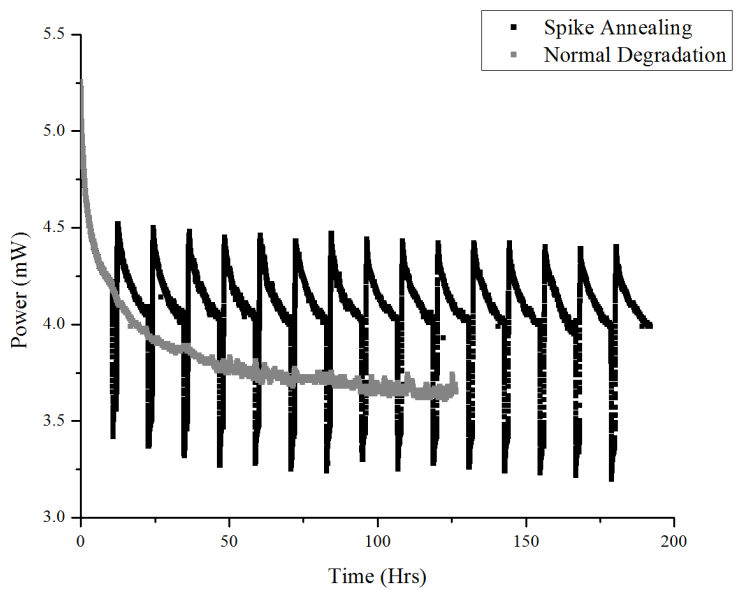

**Figure 7:** Comparison of the same 630 nm i-layer thick a-Si:H PV cell degraded at 50 °C under 1 sun (Normal Degradation) to results obtained for degradation at 50 °C under 1 sun coupled with spike annealing at 100°C on a 12 hour cycle for 192 hours (Spike Annealing).